\documentclass[10pt]{article}
\usepackage{graphicx}
\usepackage{dcolumn}
\usepackage{bm}
\usepackage{color}
\usepackage{subfigure}
\usepackage{rotating}
\usepackage{float}
\usepackage[nomarkers,figuresonly]{endfloat}
\usepackage{amsmath}
\usepackage{amssymb}

\usepackage{cite}

\usepackage{hyperref}

\usepackage{lineno}

\usepackage{microtype}
\DisableLigatures[f]{encoding = *, family = * }

\usepackage[labelfont=bf,labelsep=period,justification=raggedright]{caption}

\makeatletter
\renewcommand{\@biblabel}[1]{\quad#1.}
\makeatother

\date{}

\begin{document}

\begin{flushleft}
{\Large
\textbf{On the estimation of time dependent lift of a European Starling ($Sturnus$ $vulgaris$) during flapping flight}\\
}
%
%

Oksana Stalnov$^{1}$, 
Hadar Ben-Gida$^{2}$, 
Adam J. Kirchhefer$^{3}$,
Christoper G. Guglielmo$^{4}$,
Gregory A. Kopp$^{3}$,
Alex Liberzon$^{5}$,
Roi Gurka$^{6,\ast}$
\\
\bf{1} Faculty of Engineering and the Environment, University of Southampton, Southampton, Hampshire, SO17 1BJ, UK
\\
\bf{2} Faculty of Aerospace Engineering, Technion, Israel Institute of Technology, Haifa, 32000, Israel
\\
\bf{3} Department of Civil and Environmental Engineering, University of Western Ontario, London, ON N6A3K7, Canada
\\
\bf{4} Department of Biology, Advanced Facility for Avian Research, University of Western Ontario, London, ON, N6A5B7 Canada
\\
\bf{5} School of Mechanical Engineering, Tel Aviv University, Tel Aviv, 69978, Israel
\\
\bf{6} School of Coastal and Marine Systems Science, Coastal Carolina University, Conway, SC 29528, USA
\\
$\ast$ E-mail: Corresponding rgurka@coastal.edu
\end{flushleft}

\section*{Abstract}

We study the role of unsteady lift in the context of flapping wings in birds’ flight. Both aerodynamicists and biologists attempt to address this subject, yet it seems that the contribution of the unsteady lift still holds many open questions. The current study deals with the estimation of unsteady aerodynamic forces on a freely flying bird through analysis of wingbeat kinematics and near wake flow measurements using time resolved particle image velocimetry. The aerodynamic forces are obtained through unsteady thin airfoil theory and lift calculation using the momentum equation for viscous flows. The unsteady lift is comprised of circulatory and non-circulatory components. Both are presented over wingbeat cycles. Using long sampling data, several wingbeat cycles have been analyzed in order to cover the downstroke and upstroke phases. It appears that the lift varies over the wingbeat cycle emphasizing its contribution to the total lift and its role in power estimations. It is suggested that the circulatory lift component cannot assumed to be negligible and should be considered when estimating lift or power of birds in flapping motion.


\section*{Introduction}
Interest in the unsteady aerodynamics of flapping wing flight has been rekindled with the development of micro air vehicles (or MAVs). These MAVs fly at low Reynolds numbers, where many complex flow phenomena take place within the boundary layer. For example, separation, transition, and reattachment (of the boundary layer) can occur within a short distance along the surface of the wing and can dramatically affect the performance of the lifting surface. Despite these challenges engineers are not without prior information because nature has produced numerous examples of biological flying machines that have evolved over millions of years to efficiently fly in the low-Reynolds-number regime. One such example is the flapping flight mechanism in which the wings are not only moving forward relative to the air, but also flap up and down, bend, twist and sweep, resulting in a complicated unsteady-aerodynamic motion. Understanding the role of unsteady flapping flight will help in designing more efficient micro-flying vehicles~\cite{Bai2007}.

The flapping motion associated with the unsteady effects generally leads to enhancement of bound vortices on the lifting surface, which eventually detach, convect into the wake, and interact with other vortices~\cite{Rhozhdestvensky_Ryzhov2003}. Due to the interaction between the bound vortices on the lifting surface and the vortices in the wake, the performance of an unsteady wing is coupled with the formation and distribution of vorticity shed throughout the wing’s cycle of oscillation~\cite{Dong2006, vonEllenrieder2008}.

A useful theory to approximate unsteady aerodynamic loads is the unsteady thin airfoil theory. The roots of the theory were originally developed by Glauret~\cite{Glauret1929}, who considered simple harmonic motion. However, the complete solution of estimating the time dependent loads for a harmonically oscillating airfoil in inviscid, incompressible flow was first published by Theodorsen~\cite{Theodorsen1935}. Theodorsen's work was further complemented by von-K\'arm\'an and Sears~\cite{KarmanSears1938} who simplified the equations and presented the general unsteady thin airfoil theory. In addition to simplifying the equations, von-K\'arm\'an and Sears considered also the problem of a thin airfoil moving through a vertical gust field. The effect of unsteady inflow conditions on aerodynamic forces is considered in many applications, for example in helicopter aerodynamics~\cite{Leishman2000, Johnson2012}.

The role of unsteady forces is significant in estimating the aerodynamic performance of birds flight in flapping motion~\cite{Hedenstrom2008}. As living organisms, birds are subject to selective pressures, as such, one may assume that they operate their wings in a highly efficient manner. This notion is supported by the tendency of birds, as well as many other animals, to operate in a limited Strouhal number range between $0.2$ and $0.4$~\cite{Anderson1998, Taylor2003}. There are many factors differentiating flapping of a bird's wings from heaving or pitching of two dimensional airfoils~\cite{vonEllenrieder2008}. These differences include the presence of a body, three dimensionality of the wing and its unique motion. Ben-Gida et al.~\cite{BenGida2013} compared the formation of steady to unsteady drag at the near wake of a European Starling (Sturnus vulgaris). It was demonstrated that the unsteady drag component at the transition stages of the wingbeat phase reduce the total drag.

To model the time dependent aerodynamic forces acting on a section of a wing it is natural to start with a quasi-steady approach. The estimation of lift from the PIV measurements behind flying birds is done by applying the classical Kutta-Joukowski theorem, $L = \rho U \Gamma$, where $\rho$ is the fluid density, $U$ is the wind speed, and $\Gamma$ is the circulation calculated from the vorticity fields. Whether the lift is estimated from the Trefftz plane or from the streamwise-normal plane, it is assumed to be quasi-steady~\cite{Spedding2003, Henningsson2014}. In order to estimate the quasi-steady lift, it is sufficient to capture a portion or an entire wingbeat cycle. Former work has shown that circulation can be estimated from a single instantaneous vector map~\cite{Henningsson2008} or from synchronized velocity maps triggered to match various phases within the wingbeat cycle~\cite{Muijres2014}. Or, using several consecutive velocity maps of which a full wingbeat cycle has been reconstructed and lift was estimated for a series of velocity fields capturing the far wake behind a freely flying bird~\cite{Hedenstrom2006, Spedding2003}. Yet, incorporating the unsteady effects in lift estimations is lacking. One of the challenges in estimating the evolution of lift over time is the need to measure the wake using a technique that introduces high spatial and temporal resolution over a large period of time. In the present work, the near wake of a freely flying European starling (\textit{Sturnus Vulgaris}) has been selected as a case study of unsteady wing aerodynamics~\cite{Kirchhefer2013}. 

The aim of the present work is to evaluate the unsteady sectional lift of a flapping wing, based on experimental data acquired in the near wake of freely flying European starling ($Sturnus$ $vulgaris$) using long-duration high-speed Particle Image Velocimetry (hereafter PIV)~\cite{Taylor2010}. In the case of a flapping wing ~\cite{Zdunich2007}, the boundary layer over the wing often experiences an early transition to turbulence due to the unsteady motion and remains attached for higher angles of attack (compared to airfoil in steady condition). Such re-attachment of the boundary layer allows the use of the unsteady thin airfoil theory for lift estimation. As a first approximation the wing is estimated as rigid plate undergoing translational motion using kinematic relations~\cite{Leishman2000} based on von-K\'arm\'an and Sears~\cite{KarmanSears1938} model. Then, using high-speed PIV data, the unsteady portion of the lift is estimated over few wingbeat cycles and compared to the rigid plate lift approximation. 

\section*{Materials and Methods}
\subsection*{Theoretical modelling}
\label{sec:WakeVorticity}
The unsteady motion of a viscous fluid about a lifting surface is always accompanied with shedding of vortices into the wake. The addition point of the unsteady thin airfoil theory derivation is through the fact that the circulation around the surface is varying continuously and two-dimensional vorticity is shed off into the wake, thus allowing the use of planar wake assumption and ignoring the effect of wake deformation. In the far wake, the vortices roll-up under their own self-induced velocities to form complex wake patterns. Despite this difficulty, the contribution of the wake vortices is not significant since the influence of the shed vortices reduces with increasing distance. However, in the near wake, where the PIV measurements were conducted, the flow structures are simplified, thus allowing the use of planar wake assumption, ignoring the effects of wake deformation. Here we derive the unsteady thin airfoil theory. For clarity, we use the unsteady Bernoulli equation to evaluate the time-dependent loads. This derivation, to the best of our knowledge, has not been reported elsewhere.

\begin{figure}[p]
\centering
\includegraphics[width=0.8\textwidth]{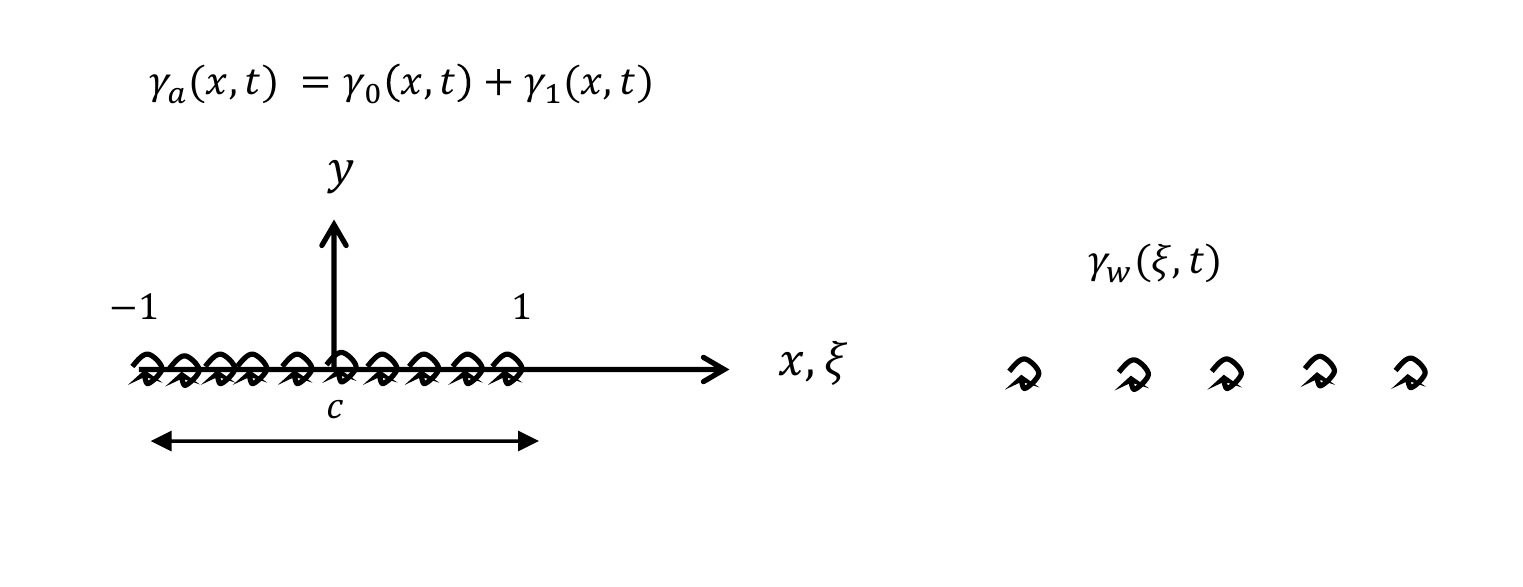}\\
\caption{{\bf Auxiliary diagram showing notation employed.}}
\label{AuxiliaryDiagram}
\end{figure}

We follow the classical definition of the unsteady thin airfoil theory ~\cite{KarmanSears1938, Theodorsen1935} notation where the chord is equal to $c=2$ (see Figure~\ref{AuxiliaryDiagram}). The leading-edge of the lifting surface is placed at $x=-1$ and trailing-edge at $x=1$, whereas the mid chord is placed at the origin of the coordinate system at $x=0$. The $y$-axis is perpendicular to the flow and the $z$-axis is in the spanwise direction of the wing.

The analysis is initiated by assuming that the vorticity distribution bound to the wing section $\gamma(x,t)$ is the sum of the quasi-steady vorticity distribution $\gamma_0(x,t)$ that would have been produced in quasi-steady motion and a wake-induced vorticity distribution $\gamma_1(x,t)$.
\begin{equation}
\gamma_a(x,t) = \gamma_0(x,t)+\gamma_1(x,t)
\end{equation}
The total circulation about the airfoil due to both the quasi-steady vorticity distribution and that induced by the wake is
\begin{equation}
    \Gamma_a=\Gamma_0+\Gamma_1
\end{equation}
where
\begin{equation}
   \Gamma_0 = \int_{-1}^1{\gamma_0(x,t)\;\mathrm{d}x}
\end{equation}
and
\begin{equation}
\Gamma_1 = \int_{-1}^1{\gamma_1(x,t)\;\mathrm{d}x}
\end{equation}
One of the fundamental assumptions in inviscid aerodynamics is that, according to Kelvin circulation theorem, the total circulation of a system is equal to zero. The total circulation about the wing section is a sum of the quasi-steady circulation $\Gamma_0$, that would be produced if the total circulation would not have been affected by the presence of the wake, and the wake-induced circulation $\Gamma_1$. When the circulation around the wing section $\Gamma_a$ is balanced with the circulation produced by the wake  $\Gamma_w$ at every time step $t$
\begin{equation}\label{TotalCirculation}
    \frac{\partial }{\partial t}\int_{-1}^1{\gamma_0(x,t)\;\mathrm{d}x}+
    \frac{\partial }{\partial t}\int_{-1}^1{\gamma_1(x,t)\;\mathrm{d}x}+
    \frac{\partial }{\partial t}\int_{1}^{\infty}{\gamma_w(\xi,t)\;\mathrm{d}\xi}=0.
\end{equation}
%
\subsubsection*{Wake-induced circulation}
\label{sec:Wake Vorticity}
The effect of the wake vortices is evaluated in accordance with the thin airfoil theory, where the Joukowski's conformal transformation is used to transform a circle in the $z'$ plane to an airfoil at the $z$ plane. The transformation relating the two planes is
\begin{equation}
    z=\frac{1}{2}
    \left (
    z^{'}+\frac{1}{z^{'}}
    \right )
\end{equation}
A single vortex element defined as $\Gamma^{'}$ at a distance of $\xi$ in the $z$-plane is located in the $z'$-plane at a distance of $\eta$. To create a unit circle which transforms into a flat plate in the $z$ plane, another vortex element with an opposite sign has to be introduced inside the unit circle at a symmetric point, which is $1/\eta$. Thus, along a unit circle the induced velocity magnitude is
\begin{equation}
    v_{\theta} = \frac{\Gamma^{'}}{2\pi|z'-\eta|}-\frac{\Gamma^{'}}{2\pi|z'-1/\eta|}
\end{equation}
which is equal to
\begin{equation}
    v_{\theta} =  \frac{\Gamma^{'}}{2\pi} \left|  \frac{\eta-\frac{1}{\eta}}{z^{'2}-z'(\eta+\frac{1}{\eta}) +1}\right |.
\end{equation}
Since $z^{'}$ is placed on the unity circle, the trigonometric identity that describes the unit circle is $z^{'}=\cos \theta + i \sin \theta$, resulting in
%

\begin{equation}
    v_{\theta} = \frac{\Gamma^{'}}{2\pi} 
    \left(
    \frac{\sqrt{\xi^2-1}}{\xi -cos\theta}
    \right)
         \label{VthLong2}
\end{equation}
 At the trailing-edge $\cos \theta=1$, thus the induced velocity at the trailing-edge is
 \begin{equation}
     v_{\theta_{TE}} = \frac{\Gamma^{'}}{2\pi}
         \sqrt{\frac{\xi+1}{\xi-1}}
\end{equation}
According to the Kutta condition, the total circulation around an airfoil is such that at any instance the flow velocity is tangential to the trailing-edge. Therefore, to meet the this condition the tangential velocity at the trailing-edge is subtracted from Eq.(\ref{VthLong2}), thus the total tangential velocity on the airfoil is
\begin{equation}
v_{\theta}=\frac{\Gamma^{'}}{2\pi}
            \left(  \frac {\sqrt{\xi^2-1}}{\xi -cos\theta}-
             \sqrt{ \frac {\xi+1}{\xi-1}} \right)=
             -\frac{\Gamma^{'}}{2\pi}
            \left(  
            \frac {1-cos \theta}{\xi -cos\theta}
            \right)
             \sqrt{ \frac {\xi+1}{\xi-1}} 
\end{equation}

The relationship between the velocity $v_{\theta}$ and the vorticity distribution over the airfoil $\gamma(x)$ according to thin airfoil theory is given by the formula $\gamma (x) = -2 v_\theta /\sin \theta$. Further, using the trigonometric relationship $x=\cos\theta$ and $\sqrt{1-x^2}=\sin\theta$, the effect of induced vorticity from a single vortex at a point $\xi$ in the wake can be written as

\begin{equation}\label{SinleIndVort}
\gamma_1^{'}(x,t)=\frac{\Gamma^{'}}{\pi(\xi-x)}\sqrt{\frac{\xi+1}{\xi-1}}\sqrt{\frac{1-x}{1+x}}
\end{equation}
where $'$ denote single vorticity element. 
The effect of the single element of vorticity $\Gamma^{'}$ can be replaced by $\gamma_w(\xi,t)\mathrm{d}\xi$.
From Eq.(\ref{SinleIndVort}) we can derive an expression for the induced vorticity of the entire wake
\begin{equation}
\gamma_1 (x,t) = \frac{1}{\pi}\sqrt{\frac{1-x}{1+x}}\int_1^{\infty} {\frac{\gamma_w(\xi,t)}{\xi-x}\sqrt{\frac{\xi+1}{\xi-1}}\;\mathrm{d}\xi}
\end{equation}
Resulting in wake-induced circulation
\begin{equation}\label{Gamma1}
\Gamma_1  = \int_1^{\infty} {\gamma_w(\xi,t) \left (\sqrt{\frac{\xi+1}{\xi-1}}-1 \right) \;\mathrm{d}\xi}
\end{equation}

\subsubsection*{Unsteady thin airfoil theory}\label{sec:UnSteadyTheory}
K\'arm\'an and Sears~\cite{KarmanSears1938} applied the principle that, in accordance with the Newton's second law, the product of density and the rate of change of the total momentum at any instance is equal to the lift. In the current study, an estimation of the lift due to an unsteady motion is based on the integration of normal pressure difference along the chord
\begin{equation}
    L(x,t)=\int_{-1}^{1} {\Delta p(x,t) \;\mathrm{d}x}.
\end{equation}
The pressure difference $\Delta p (x,t)$ in terms of the chordwise vorticity distribution $\gamma_a (x,t)$ is expressed by the unsteady Bernoulli equation~\cite{KatzPlotkin}
\begin{equation}
    \Delta p (x,t) = \rho U \gamma_a (x,t) +  \rho \frac{\partial}{\partial  t}\int_{-1}^x{\gamma_a}(x_0,t)\;\mathrm{d}x_0.
\end{equation}\label{press diff}
Thus, the lift can be written as
\begin{equation}
\label{InsteadyLift21}
    L(x,t)= \rho U \int_{-1}^{1} {\gamma_a (x,t)\;\mathrm{d}x} +
            \rho \int_{-1}^{1} \frac{\partial}{\partial t}\int_{-1}^x{\gamma_a}(x_0,t)\;\mathrm{d}x_0 \;\mathrm{d}x.
\end{equation}
Integration by parts of the second integral on the right hand side of Eq.(\ref{InsteadyLift21}) yields

\begin{equation}
\begin{split}
    \rho \int_{-1}^{1} \frac{\partial}{\partial t} \int_{-1}^{x}{\gamma_a}(x_0,t)&\;\mathrm{d}x_0 \;\mathrm{d}x = \\
    &\rho \frac{\partial}{\partial t} 
    \left ( 
    \left [ x \int_{-1}^{x} \gamma_a (x_0,t) \;\mathrm{d}x_0 \right] \,\Big|_{-1}^{1}
    -\int_{-1}^{1} \gamma_a(x,t) x \;\mathrm{d}x 
    \right )
    \end{split}
\end{equation}
where it is recognized that
\begin{equation}
   \left[ x \int_{-1}^{x} \gamma_a(x_0,t) \;\mathrm{d}x_0\, \right] \Big|_{-1}^{1} =  \int_{-1}^{1} \gamma_a(x,t)\;\mathrm{d}x.
\end{equation}
The lift terms in Eq.(\ref{InsteadyLift21}) can then be rearranged as
\begin{equation}
\label{UnsteadyL}
\begin{split}
L = & \rho U \int_{-1}^{1}\gamma_0(x,t) dx
     -\rho \frac{\partial}{\partial t} \int_{-1}^{1} {x\gamma_0(x,t) \;\mathrm{d}x}
     +\rho \frac{\partial}{\partial t} \int_{-1}^{1} {\gamma_0(x,t)  \;\mathrm{d}x}\\
    & + \rho U \int_{-1}^{1} {\gamma_1(x,t) \;\mathrm{d}x}
    - \rho \frac{\partial}{\partial t} \int_{-1}^{1} x\gamma_1 (x,t) \;\mathrm{d}x
      +\rho \frac{\partial}{\partial t} \int_{-1}^{1} \gamma_1 (x,t) \;\mathrm{d}x.
\end{split}
\end{equation}
The first term in Eq.(\ref{UnsteadyL})
\begin{equation}
    L_0=\rho U \int_{-1}^{1}{\gamma_0(x,t) \;\mathrm{d}x}
\end{equation}
is the quasi-steady Kutta-Joukowski lift component. In permanently maintained flow conditions this would be the only lift component. In unsteady flow conditions the quasi-steady lift component only partially contributes to the total lift and it is determined by evaluating instantaneous angle of attack.
The second term in Eq.(\ref{UnsteadyL})
\begin{equation}
\label{L_1_KS}
    L_1 = -\rho \frac{\partial}{\partial t} \int_{-1}^{1}{x\gamma_0(x,t)\;\mathrm{d}x}
\end{equation}
is the apparent (or added mass) lift component that accounts for the reaction due to the mass of fluid directly accelerated by the wing. Following the von-K\'arm\'an and Sears discussion on time derivative, the fifth term of Eq.~(\ref{UnsteadyL}) is equal to
\begin{equation}
\begin{split}
    \rho \frac{\partial }{\partial t}&\int_{-1}^{1} {x\gamma_1(x,t) \;\mathrm{d}x} = \rho \frac{\partial }{\partial t}\int_{1}^{\infty} \gamma_w(\xi,t) \;\mathrm{d}\xi+
    \rho \frac{\partial }{\partial t} \int_{1}^{\infty}({\sqrt{\xi^2-1}-\xi) \gamma_w(\xi,t) \;\mathrm{d}\xi}\\
    &=\rho \frac{\partial }{\partial t}\int_{1}^{\infty} \gamma_w(\xi,t) \;\mathrm{d}\xi
    +\rho U \int_{1}^{\infty}{\left ( \sqrt{\frac{\xi+1}{\xi-1}}-1-\frac{1}{\sqrt{\xi^2-1}} \right )\gamma_w(\xi,t)\;\mathrm{d}\xi}\\
    &=\rho \frac{\partial }{\partial t}\int_{1}^{\infty} \gamma_w(\xi,t) \;\mathrm{d}\xi
    +\rho U \int_{-1}^{1}\gamma_1(\xi,t)\;\mathrm{d}\xi -\rho U \int_{1}^{\infty}{\frac{\gamma_w(\xi,t)}{\sqrt{\xi^2-1}} \;\mathrm{d}\xi}\\
\end{split}
\label{TimeDerivative}
\end{equation}
It should be noted, that the third term of equation  Eq.~(\ref{TimeDerivative}) is  the wake-induced lift
\begin{equation}
L_2 = \rho U \int_{1}^{\infty}{\frac{\gamma_w(\xi,t)}{\sqrt{\xi^2-1}} \;\mathrm{d}\xi}.
\end{equation}

Evaluation of $L_2$ term requires either an assumption about the unsteady motion, keeping track of the shed vorticity into the wake, or including the shed vorticity through a convolution integral.
The forth term in Eq.(\ref{UnsteadyL}) is cancelled with the second term of equation Eq.(\ref{TimeDerivative}). Adding the third and sixth terms in Eq.(\ref{UnsteadyL}) with the first term in Eq.(\ref{TimeDerivative})
results in the Kelvin theorem (see Eq.(\ref{TotalCirculation})), therefore, this summation is zero. These leads to the time dependent lift (Eq.(\ref{UnsteadyL})) that can be written as the sum of three terms


\begin{equation}
\label{Sears_Lift_all}
L=\rho U \Gamma_0-\rho\frac{\partial}{\partial t} \int_{-1}^1 {x\gamma_0(x,t) \;\mathrm{d}x}+
        \rho U\int_1^{\infty} \frac{\gamma_w (\xi,t)}{\sqrt{\xi^2-1} }\;\mathrm{d}\xi.
\end{equation}
Eq.(\ref{Sears_Lift_all}) is the result of von-K\'arm\'an and Sears~\cite{KarmanSears1938}. The first term is the quasi-steady lift $L_0$ produced by the bound vorticity. The second term $L_1$ represents the apparent (added) mass contribution to lift component and it results from the inertia of the fluid moving with the lifting surface. The third term $L_2$ is the induced lift component that produced by the wake vorticity. It should be noted that the contribution to the lift due to wake-induced vorticity $\gamma_1(x,t)$ is cancelled out in the derivation of the equations.

This result coincides with Theodorsen~\cite{Theodorsen1935}, who suggested to divide the time dependent lift into circulatory and non-circulatory components, namely $L_C$ and $L_{NC}$, respectively. The non-circulatory lift can be referred as the virtual mass effect or the acceleration reaction term~\cite{Batchelor1967} and the circulatory lift component is generated from the vortical flow around the lifting surface. We should point out that the non-circulatory lift term, $L_{NC}$, which is related to the added mass lift, is identical to the $L_1$ term presented by von-K\'arm\'an and Sears~\cite{KarmanSears1938} (see Eq.(\ref{L_1_KS})), i.e. $L_{NC}=L_1$. The circulatory lift term $L_C$ is equal to the sum of the two remaining lift components, these are the quasi-steady lift and the wake-induced lift, i.e. $L_C=L_0+L_2$. 

\subsection*{Experimental Apparatus}
\subsubsection*{Wind tunnel}
The experiments reported herein conducted in a hypobaric climatic wind tunnel at the Advanced Facility for Avian Research (AFAR) at the University of Western Ontario. A detailed description of the wind tunnel, the experimental technique and the bird is given by Ben-Gida et al.~\cite{BenGida2013} and Kirchhefer et al.~\cite{Kirchhefer2013}. Herein we provide a short description, for brevity. The wind tunnel is closed loop type with an octagonal test section. The cross-sectional area is 1.2 m$^2$, preceded by a $2.5:1$ contraction ratio. The width, height and length of the test section are $1.5\,$m, $1\,$m, and $2\,$m, respectively. The control of speed, pressure, temperature, and humidity in the wind tunnel enables to simulate flight conditions at high altitudes as experienced by birds during long distanced migratory conditions. The bird is introduced into the test section through a partition that is located between the downstream end of the test section and the diffuser. The turbulence intensity measured by the hot-wires was lower than 0.3\% over the entire test section with a uniformity of 0.5\%. A fine size net was placed at the upstream end of the test section to prevent the bird from entering the contraction area. The flight conditions reported in this work correspond to atmospheric static pressure, a temperature of $15^{\circ}$C, and relative humidity of 80\%.
\begin{figure}[p]
    \centering
    \includegraphics[width=0.5\textwidth] {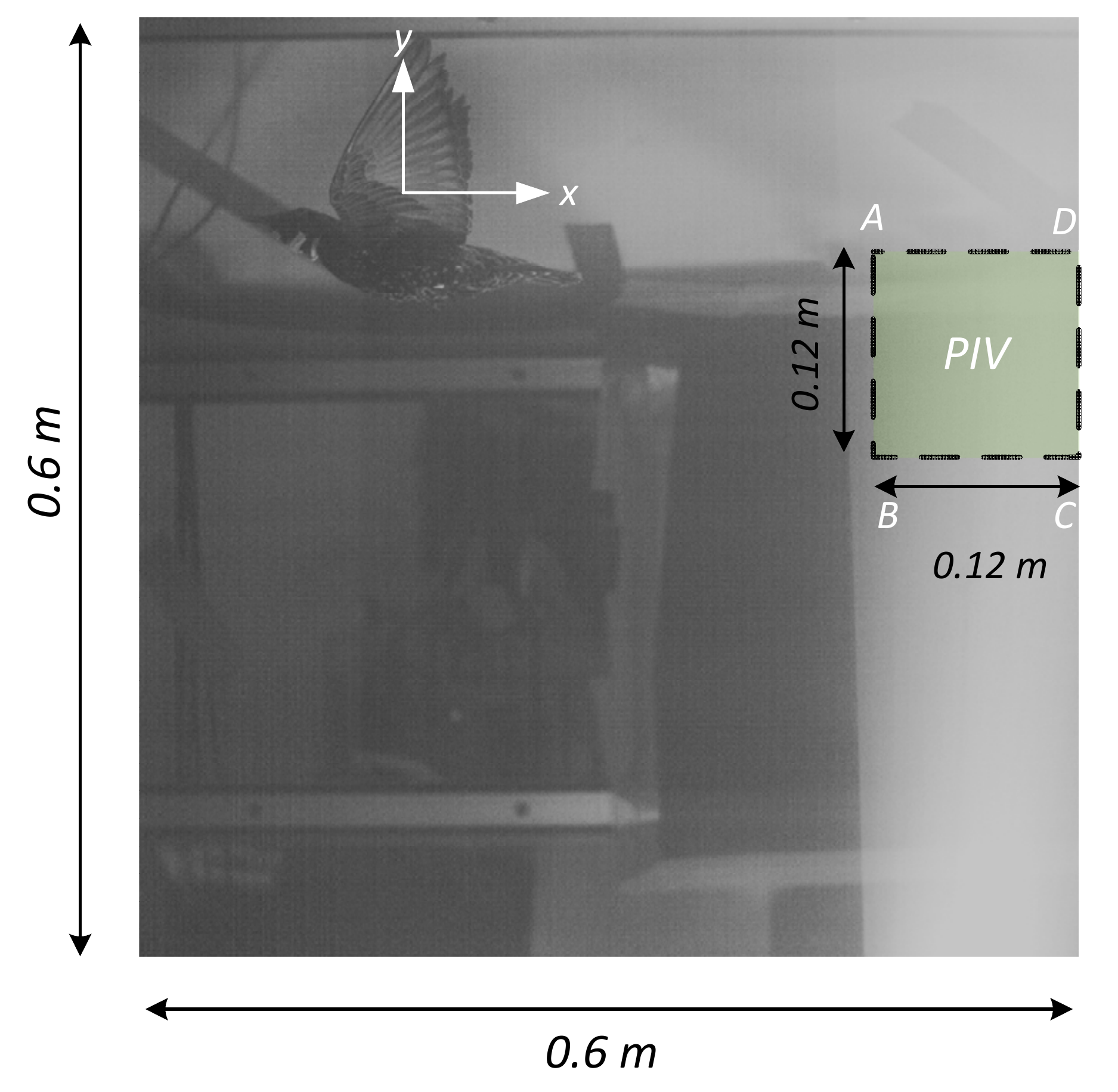}\\
    \caption{{\bf The large image shows the kinematic camera field of view and the small window marked `PIV' is the PIV camera field of view.}}
    \label{fig:starling}
\end{figure}

\subsubsection*{The Bird - European Starling}
The wake measurements (as illustrated in Figure~\ref{fig:starling}) were sampled from a European starling ($Sturnus$ $vulgaris$) that was trained to fly in the AFAR wind tunnel. The bird's wings had an average chord of $c=6\,$cm, a maximum wingspan of $b=38.2\,$cm ($b_{semi}=19.1\,$cm) and an aspect ratio (wingspan squared divided by the wings’ lifting area) of 6.4. The wind speed was set to $U_{\infty}=12\,$m/s. The wingbeat frequency, $f$, was $13.3\,$Hz on average, and the average peak-to-peak wingtip vertical amplitude, $A$, was $28\,$cm. These quantities correspond to a chord-based Reynolds number of $4.8 \times 10^4$, a Strouhal number, $St=Af/U_{\infty}=0.3$, and a reduced frequency, $k=\pi fc/U_{\infty}=0.2$. The bird’s mass was $78\,$g and a lateral body width of $4\,$cm. Specially designed safety goggles (Yamamoto Cogaku Co. model  YL600) were adjusted to the bird while flying at the tunnel. In addition, a collection of optoisolators operated by six infrared transceivers were integrated into the PIV system in order to prevent direct contact between the bird and the laser sheet. The optoisolators triggered the laser only when the bird was flying upstream further from the PIV field of view. All animal care and procedures were approved by the University of Western Ontario Animal Use Sub-Committee (protocols 2006-011, 2010-216). 

\subsubsection*{Long duration time resolved PIV}
Flow measurements were taken using the long-duration time-resolved PIV system developed by Taylor et al.~\cite{Taylor2003}. The PIV system consists of a $80\,$W double-head diode-pumped Q-switched Nd:YLF laser at a wavelength of $527\,$nm and two CMOS cameras (Photron FASTCAM-1024PCI) with spatial resolution of $1024 \times 1024$pixel$^2$ at a sampling rate of $1000\,$Hz. The PIV system is capable of acquiring image pairs at $500\,$Hz using two cameras for a continuous period of 20 minutes. Olive oil aerosol particles, $1\,\mu$m in size on average~\cite{Echols_Young1963} were introduced into the wind tunnel using a Laskin nozzle from the downstream end of the test section so that it did not cause a disturbance to the flow in the test section or to the bird. The system is designed to work either in a 2D or Stereo mode. Herein, we used one camera for the PIV whilst the other camera was used for measuring the wingbeat kinematics simultaneously with the PIV. The PIV camera's field of view was $12  \times 12$cm$^2$ corresponding to $2c$ by $2c$. The velocity fields were computed using OpenPIV ~\cite{Taylor2010} using $32 \time 32$pixel$^2$ interrogation windows with 50\% overlap, yielding a spatial resolution of 32 vectors per average chord, equal to 1.8 vectors per millimetre. In the current experiments, 4,600 vector maps were recorded, and out of this dataset 650 vector maps contained features of the near wake behind the starling's wing. The measured wake was located 4 wing chord lengths behind the right wing. The wake was sampled in the streamwise-normal plane at $2\,$ms intervals ($500\,$Hz), therefore, both the downstroke and the upstroke phases were temporally resolved.

\subsubsection*{Kinematic measurements}
To relate the wake measurements to the kinematic motion of the bird's wings, an analysis of the kinematic motion has been undertaken. The simultaneous measurements also enables a point of comparison between the estimation of lift through the unsteady thin airfoil theory and lift calculation using the momentum equation for viscous flows. The field of view by the CMOS camera is $9c$ by $9c$ (corresponding to 54 x 54cm$^2$). Figure~\ref{fig:starling} depicts a sample image of the Starling flying in the wind-tunnel as captured by the camera. The box marked with the `PIV' label indicates the location of the measured velocity fields from the PIV system. In addition, a floor-mounted camera operating at $60\,$Hz was used to record the spanwise position of the bird with respect to the laser sheet illumination. Therefore, these images provided the identification of the measured PIV plane and its location in respect to the wing; so that, the wake velocity field associated with the spanwise location across the wing is $0.15b/2$ from the root. The floor-mounted camera was not synchronized with the PIV or the kinematic measurements, so the two time histories were synchronized manually based on the presence of the laser light in the images. Once synchronized, spanwise positions were assigned to the wake data acquired at $500\,$Hz based on interpolation from the simultaneously recorded spanwise positions.  

\subsubsection*{Error estimation}
An error analysis based on the root sum of squares method has been applied to the velocity data and the wing kinematics. The errors were estimated as: 2.5\% for the instantaneous velocity values, 12\% for the instantaneous vorticity and 3\% for the lift values~\cite{Raffel2007}. The error introduced in the kinematic analysis resulted from the spatial resolution of the image and the lens distortion leading to an estimated error of 5\% in the wing displacements. 

\section*{Results and Discussions}
\label{Results}

The PIV flow field measurements and the bird's kinematics were each analysed separately in order to estimate the time dependent component of the lift generated by the flapping motion of the starling using the inviscid and the viscid approaches. Linear lift theory (see the theoretical modelling section) was used to estimate the lift from the bird's kinematics, whereas a viscous flow theory, derived by Wu ~\cite{Wu1981} and applied by Panda~\cite{Panda1994}, was implemented to estimate the lift from the near wake flow fields measured by the PIV. The present work considers a comparison between the linear theories~\cite{Theodorsen1935, KarmanSears1938} and the viscous flow theory~\cite{Wu1981}, for the lift generated by the starling. Both approaches will emphasize the role of the time dependent lift components.

\begin{figure}[p]
\centering
\subfigure[wingbeat 1-- upstroke]
{\includegraphics[width=0.3\textwidth]
{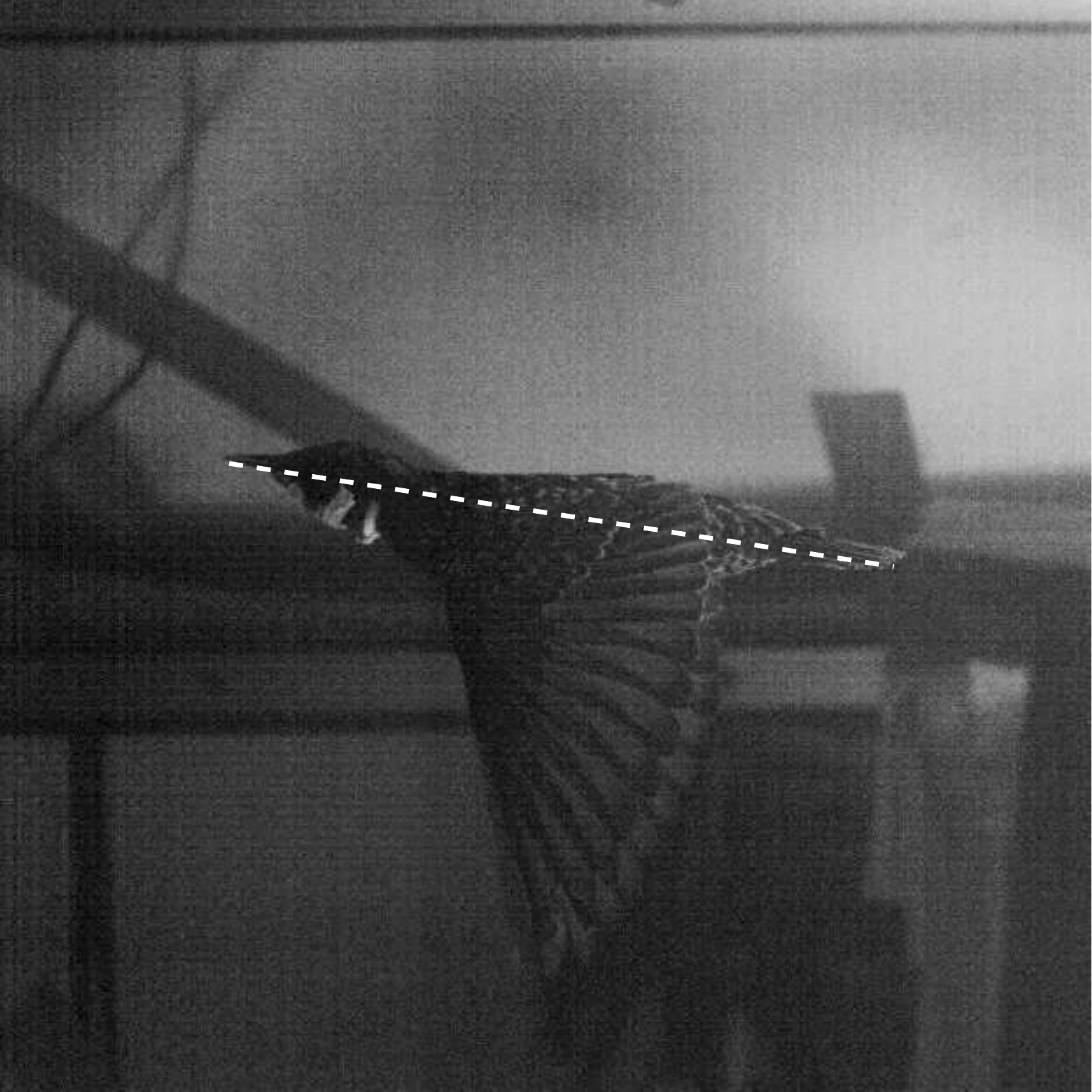}}
\subfigure[wingbeat 1--downstroke]
 {\includegraphics[width=0.3\textwidth]
{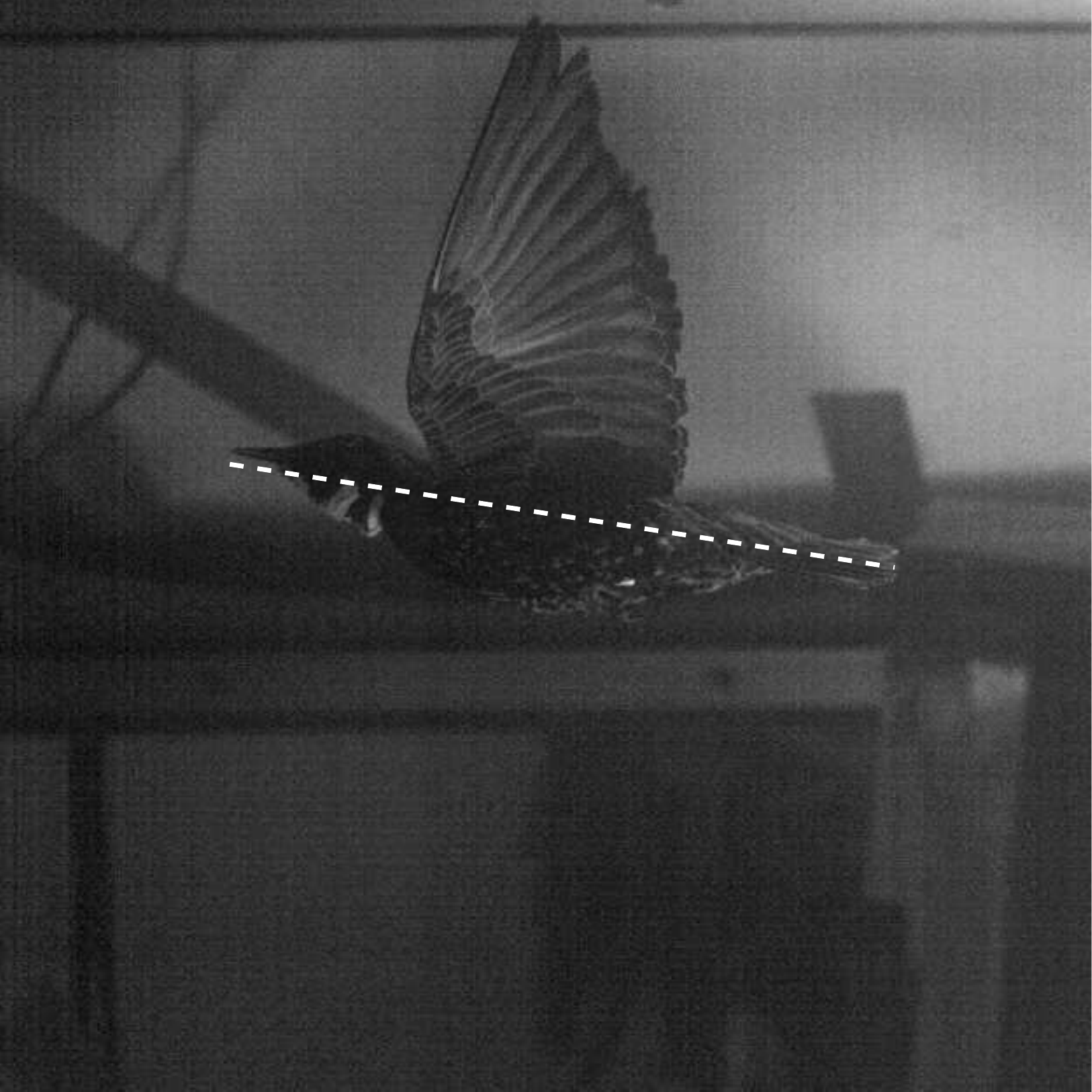}}\hspace*{5mm}
\\
\subfigure[wingbeat 2--upstroke]
{\includegraphics[width=0.3\textwidth]
{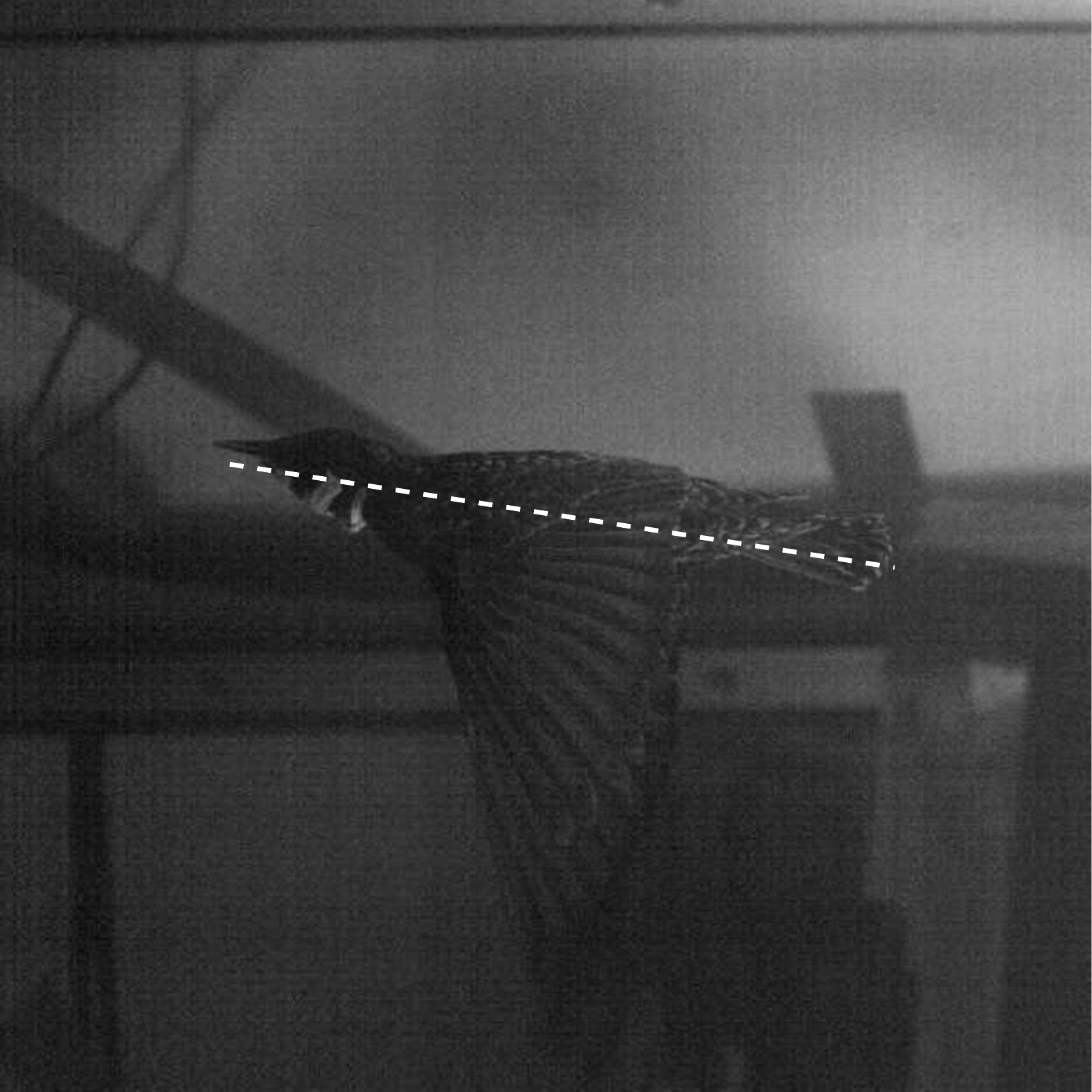}}
\subfigure[wingbeat 2--downstroke]
 {\includegraphics[width=0.3\textwidth]
{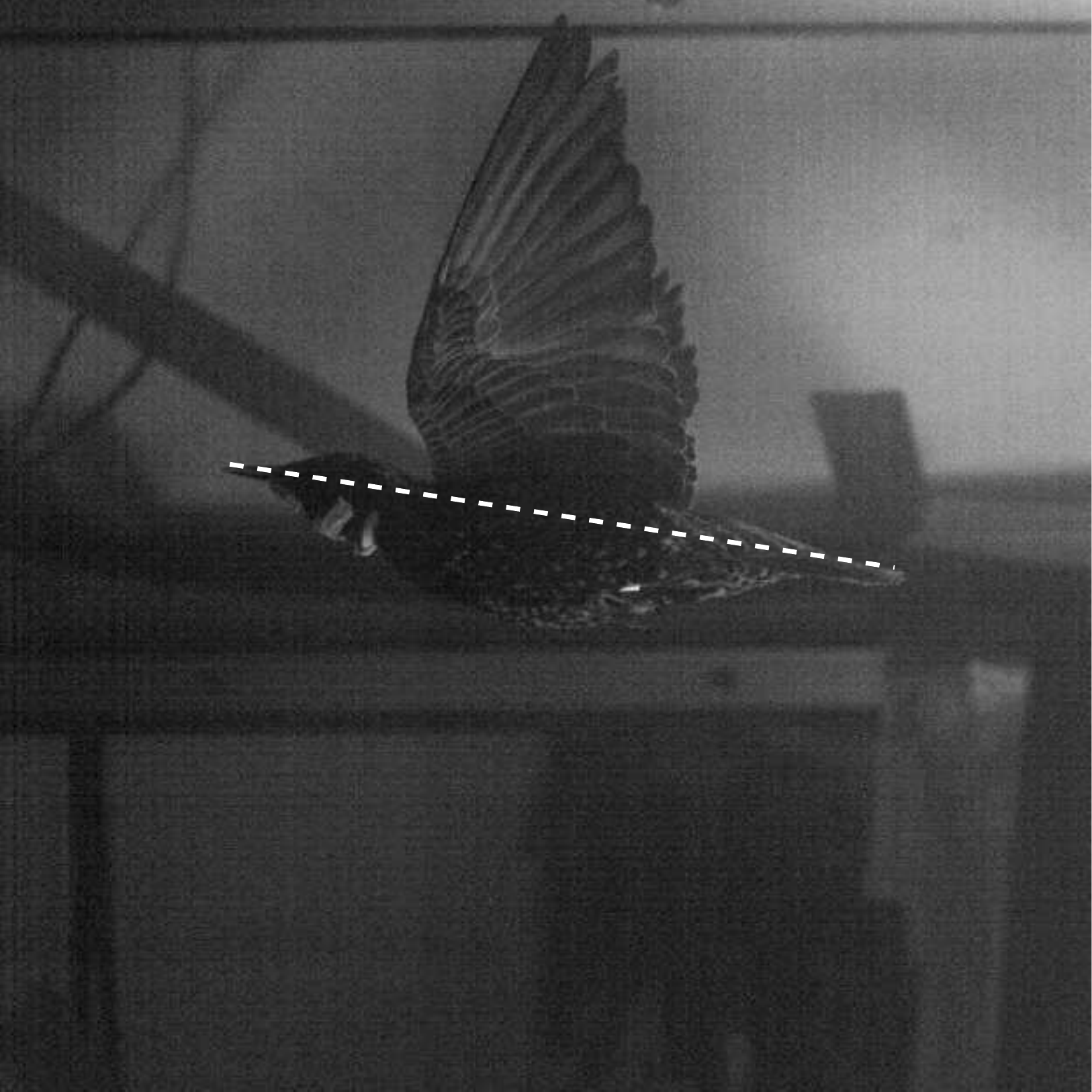}}\hspace*{5mm}
\\
\subfigure[wingbeat 3--upstroke]
{\includegraphics[width=0.3\textwidth]
{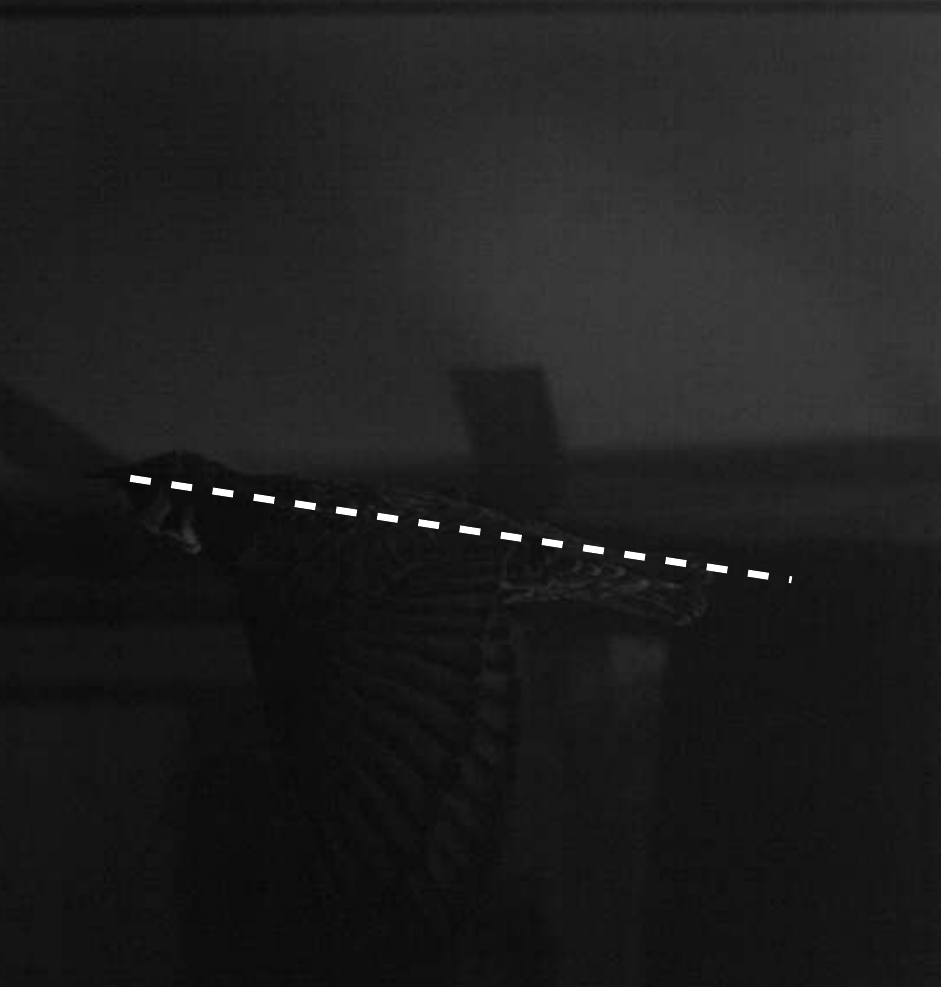}}
\subfigure[wingbeat 3--downstroke]
 {\includegraphics[width=0.3\textwidth]
{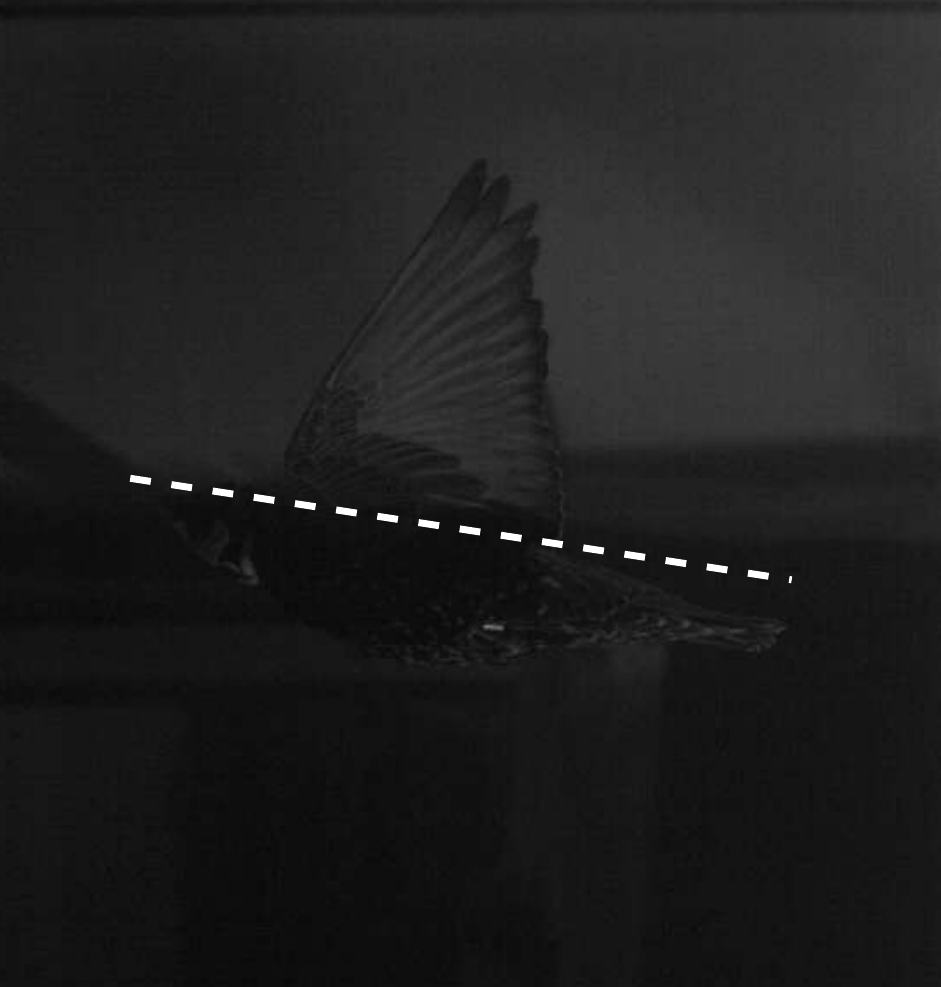}}\hspace*{5mm}
\\
\caption{{\bf Side view of the European Starling at the wind tunnel.} The white line is placed to provide spatial reference of the bird's body. The white line is inclined at  $8.8^{\circ}$ with the free-stream velocity. The left and right images correspond to beginning of the downstroke and upstroke, respectively.}
\label{f:Kinematics}
\end{figure}

The bird kinematics and the near wake velocity field were captured simultaneously while the starling was flying in the tunnel and flapping its wings continuously. During flapping, birds generate lift and thrust. The lift is necessary to support the bird's body weight and the thrust is required to overcome the drag. The data presented herein correspond to no-maneuver and no-acceleration conditions. The spatial location of the bird's body at the beginning of the downstroke and upstroke phases of flight are shown in Figure~\ref{f:Kinematics} for three consecutive wingbeats. It can be seen that the bird does not accelerate in the streamwise or vertical directions~\cite{BenGida2013}. Hence, the following kinematic analysis can be performed assuming negligible acceleration. 

%
\subsection*{Estimation of time-dependent lift from the bird kinematics} 

As presented in the theoretical modelling section, the linear theories were derived within the framework of potential theory, which assumes inviscid fluid with small disturbances and a plane wake. One can use such theories to estimate the time dependent lift components from the kinematics of a wing section with relatively good precision~\cite{Leishman2000}. We choose to use the guidelines provided by Leishman~\cite{Leishman2000} and estimate the time dependent lift from the bird's kinematics.

Using the aforesaid unsteady thin airfoil theory, we estimate the lift generated by the flapping wings motion of the European starling, as captured through the wings kinematic images. For simplicity, we describe the kinematics of a flapping wing by a pure two-dimensional plunging motion, which involves a heaving up and down of the wing section that results in a variation of the effective angle of attack~\cite{Leishman2000}. In such motion the variation of the vertical displacement with time can be described as follows

\begin{figure}[p]
\centering
  \includegraphics[width=0.3\textwidth]
  {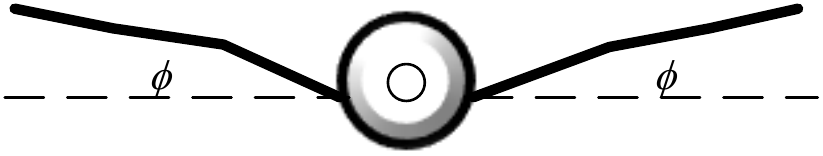}\\
    \caption{{\bf Auxiliary diagram of the flapping motion with the definition of the flapping angle.} View in the direction of the flow.}
    \label{FlappingAngle}
\end{figure}
\begin{equation}
    h_f(t)=h_{f_0} \cos( \omega t)
\end{equation}
where $h_{f_0}$ is the plunging motion amplitude and $\omega$ is the angular velocity. For a flapping wing the plunging amplitude $h_{f_0}$ is a function of the distance from the shoulder joint. Assuming the elastic deformation of the wing is negligible we can describe the plunging amplitude as linearly increasing function towards the wing tip. Thus, the vertical displacement amplitude at spanwise distance $\eta=0.15b/2$ from the wing root can be reduced to $h_{f_0}=\eta \cos (\phi)$, where $\phi$  is the wing tip angle, as depicted in Figure~\ref{FlappingAngle}. Kinematic images of the starling~\cite{Kirchhefer2013} depicted the wings as they oscillate in a periodic manner, where the range of the angular positions is $-55^{\circ}<\phi<19^{\circ}$

Due to the unique bone and muscle structure of the bird's wing, during flapping flight the inner part of the wing experiences less twisting motion than the outer part which accounts for most of the thrust production~\cite{Dhawan1991}. Hence, the variation of the local angles of attack at the inner part of the wing is small compared to those at the outer part. Therefore, we can describe the effective angle of attack as a result of the horizontal free-stream velocity and the vertical velocity component due to plunging motion, $ \alpha_{e} = \tan^{-1}(\dot{h}(t)/U_{\infty})$. By assuming small angles of attack we can simplify the effective angle of attack to $ \alpha_{e} \approx (\dot{h}(t)/U_{\infty})$. Consequently, the quasi-steady lift component can be estimated by the thin airfoil theory~\cite{Anderson1985}, where the non-dimensional lift coefficient is a function of effective angle of attack and the corresponding lift component is equal to
\begin{equation}
L_{0}=\pi\rho U^2 c \left[ \frac{\dot{h}(t)}{U}  \right]
 \label{TH_L0}
\end{equation}

Following the unsteady thin airfoil theory~\cite{KarmanSears1938, Theodorsen1935}, the added-mass lift component is a result of flow acceleration, and thus arises from the unsteady term in the Bernoulli equation that accounts for the pressure force required to accelerate the fluid in the the vicinity of the wing. For the wing section moving normal to its surface at velocity $v(t)$, the non-circulatory fluid force acting on the surface is equal to the product of apparent mass and acceleration. Thus, a body moving in an unsteady motion must overcome acceleration in addition to its own inertial force. Therefore, the apparent mass (or non-circulatory~\cite{Theodorsen1935}) lift component can be estimated from the kinematic motion accordingly
\begin{equation}
    L_1=\pi \rho U^2 \frac{c^2}{4} \left[ \frac{\ddot{h}(t)}{U^2}\right]
     \label{TH_L1}
\end{equation}
where $\dot{h}(t)$ is the time derivative of vertical displacement~\cite{Leishman2000}. 

Here, the effect of the wake-induced lift component $L_2$ is determined by assuming harmonic motion~\cite{Theodorsen1935} at a frequency $\omega$, yielding $\gamma_w(\xi,t) = g e^{i \omega (t-\xi/U)}$. Using Theodorsen's function $C(k)$, which accounts for the effect of the shed vortices on the unsteady aerodynamic loads, we can calculate the wake-induced lift component as follows
\begin{equation}
    L_2(t) = (C(k)-1)L_0
    \label{TH_L2}
\end{equation}
where the definition of Theodorsen's function is~\cite{Theodorsen1935}
\begin{equation}
    C(k) = \frac{H_1^{(2)}(k)}{H_1^{(2)}(k)+i H_0^{(2)}(k)}.
\end{equation}
Here $H_n^{(2)}=J_{\nu}-iY_{\nu}$ is the Hankel function of the reduced frequency $k$, where $J_{\nu}$ and $Y_{\nu}$ are Bessel functions of the first and second kind, respectively. The lift reduction function $C(k)$ falls gradually to value of $0.5$ as $k$ goes to infinity. The effect of $C(k)$ as producer of phase lag takes over very quickly, where the maximum rotation of the vector occurs around $k=0.2$. It should be noted that the reduced frequency values used by many small passerines, such as the European starling, lays well in the range of $0.1<k<0.3$ ~\cite{Shyy2008}. Apparently, these small passerines fly at the region where the effect of the lift reduction function is the strongest.

\begin{figure}[p]
  \centering
  \includegraphics[width=1\textwidth]{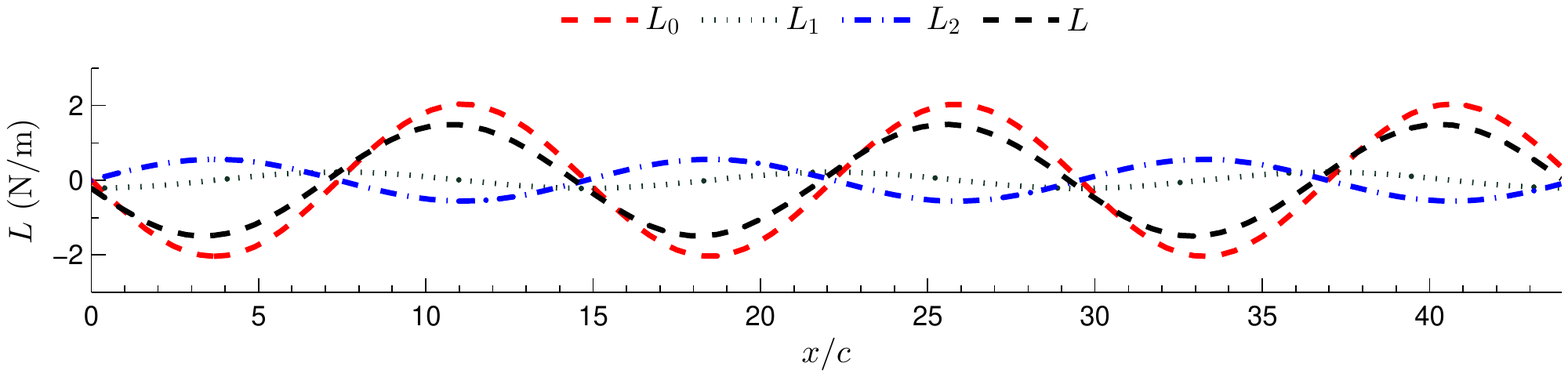}\\
  \caption{{\bf Time dependant components of unsteady lift terms, as estimated from the wing kinematics.} The terms $L_0$,  $L_1$,  $L_2$, are estimated from Eqs.~\ref{TH_L0},~\ref{TH_L1},~\ref{TH_L2}, respectively. The term $L=L_0+L_1+L_2$ is the total lift.}
  \label{WKLE}
\end{figure}
Figure~\ref{WKLE} shows the time variation of the three lift components ($L_0$, $L_1$ and $L_2$). The sum of the three components correspond to the total time dependent lift $L$ generated during the flapping motion. It appears that the non-circulatory (or added mass) contribution to lift is the smallest among the three components. Yet, this contribution is not negligible and in fact is equal to about half of the the lift generated by the induced-wake. Both components are significantly smaller compared to the quasi-steady lift component. According to the harmonic assumption the wake-induced lift is in anti-phase to the quasi-steady lift component. The generation of the circulatory lift components comprises of two terms that in fact are counter to each other. The theory explains the negative work done by the induced vorticity during the upstroke phase as the wing approaches maximum value of the lift. Rival et al.~\cite{Rival2009} utilized similar principles to study the effect of leading-edge vortex on the formation of dynamic stall over an airfoil. 

\subsection*{Estimation of the circulatory time-dependent lift component from the bird's near wake vorticity field}

The unsteady thin airfoil theory is a useful tool that provides a good approximation of the time dependent aerodynamic loads. Nevertheless, the theory, which bounds to two-dimensional inviscid flows, underestimates loads in complex flows where viscosity cannot be ignored. Therefore, another approach is needed. One of the methods for estimating aerodynamic loads is based on the flow field at the wake region. PIV provides high resolution spatial data with sufficient accuracy that enables the estimation of such loads from the wake of bluff bodies~\cite{Raffel2007}. However, despite many advances in the current state-of-the-art in experimental diagnostics, practical application of PIV to estimate time dependent forces from wake flow-field measurements are challenging. These efforts are limited to 2D planes. This limitation is mainly due to the fact that the 3D flow-field measurements are restricted by relatively small volume size and low Reynolds number flows. Thus, the most practical approach to estimate the time dependent lift component is from 2D plane measurements. 

Within the avian research community, the most common 2D approach is concerned with measurements of the vorticity field in the far wake Trefftz plane and application of the Kutta-Joukowski quasi-steady theorem in order to estimate the lift. This approach is based on the classical assumption that the vortex lines behind a lifting surface roll-up when they propagate downstream into the wake, and they bundle into tip vortices. Thus, the far wake is dominated by the tip vortices. This approach is appealing as the entire wake structures can be captured by a single plane, provided these measurements are acquired in the far wake. However, measurements conducted in the Trefftz plane are highly inaccurate due to plane normal velocity component and may lead to significant errors that are hard to ignore~\cite{Waldman2012}. Furthermore, these measurements allow estimation of only the mean quasi-steady total lift and not the time dependent evolution of the lift.

The second 2D approach is concerned with measurements of the flow field in the streamwise plane at the near or far wake. A brief summery of the PIV measurement acquired in the wake of freely flying birds can be found in figure 1 at Kirchhefer et al. 2013~\cite{Kirchhefer2013}. Amongst which are flow measurements in the wake of Thrush Nightingale~\cite{Spedding2003}, Robin~\cite{Hedenstrom2006}, Swift~\cite{Henningsson2008}  and bats~\cite{Hubel2009, Muijres2011}). PIV measurements in the streamwise plane are considered to be accurate, with some errors related to the spanwise velocity component~\cite{Raffel2007}. As it has been indicated previously, one of the first applications of PIV technique to estimate the lift from the wake of freely flying bird is attributed to Spedding et. al.~\cite{Spedding2003}. In this work, the lift was estimated based on quasi-steady Kutta-Joukowski thin airfoil theory. This simplified approach, which has been followed by many other researchers, neglects the effects of added mass and wake-induced vorticity on the time dependent lift components. 

In the current work, the near wake flow-fields were captured simultaneously to the bird's kinematic motion, shown in Figure~\ref{f:Kinematics} and discussed in the previous section, thus, allowing one to relate the wake flow-field structures to the bird's kinematic motion. The PIV measurements were taken at the inner part of the right wing (from the bird's perspective), at a spanwise distance of $z = 0.15b/2$ from the wing root. In order to shed light on the wake structures that manifest the bird's lift, a visualization of the entire wingbeat during a single flapping cycle is performed by generating a wake composite image from multiple PIV realizations. A similar approach was first applied by Spedding et al.~\cite{Spedding2003} in which PIV measurements (from separate wingbeat cycles) were arranged to represent a complete and representative wavelengths of the wake. 

The wake composite is formed by plotting sequential PIV realizations, each image is offset to one-another in the streamwise direction. The offset of the PIV images is calculated as $U_{\infty}\Delta t \cdot n$. The generation of a wake composite provides a useful tool for observing the time-series of measurements representing the wake of a wingbeat cycle. The procedure was performed using the PIV flow-fields collected at a sampling rate of $500\,$Hz that is significantly higher than the bird's flapping frequency of $13\,$Hz. Therefore, a pattern of vorticity appearing in one frame also appears in the consecutive frame –- only phase-shifted. The wake structures that appear `downstream' in the wake composite image happen earlier in time, while the structures that appear `upstream' in the composite actually happen later in time. In a sense, the generation of the wake composite image invokes Taylor’s hypothesis~\cite{Taylor1938} in which the characteristics of the flow are advected through the field of view, where the offset of one image to the next is based on the free stream speed. It should be noted that the typical offset of $U_{\infty}\Delta t \cdot n$ between images is $\sim 0.4c$ and an instantaneous PIV measurement has a spatial dimension of $2c$. Therefore, at any location in the wake composite image, there are several overlapping images that can be used to ascertain the instantaneous wake characteristics over the streamwise distance of $2c$ to compare with the wake composite at the same location. 

The wake features are shown through fluctuating velocity and vorticity fields, where the spanwise vorticity is defined as follows
\begin{equation}
 \omega_z (t) = \frac{\partial v}{\partial x}-\frac{\partial u}{\partial y}
 \end{equation}
and is evaluated directly from the PIV flow fields using a least squares differentiation scheme. Here $u$ and $v$ are streamwise and transverse velocity components, respectively. 

Estimation of time dependent lift generated during the flapping motion of the starling is evaluated from the near wake velocity maps by utilizing the viscid approach derived by Wu~\cite{Wu1981} based on the Navier-Stokes equation~\cite{Batchelor1967, Lamb1945}. The generalized formulation that conveniently describes the aerodynamic forces exerted by a fluid on a solid body immersed in, and moving relative to the fluid, is equal to inertial force due to the mass displaced by the solid body and a term proportional to the time of change of the first moment of the vorticity field~\cite{Wu1981, Lighthill1986}, as follows:
\begin{equation}
L(t)=-\rho \frac{d}{dt}
\left[\iint {x\omega_z(t) \; dx dy}\right]
+m'\frac{dU}{dt}
\label{Wu_Lift_General}
\end{equation}
where $m'$ is the mass of the fluid displaced by the solid body. One can immediately recognize that the second term in Eq.(\ref{Wu_Lift_General}) is the added mass lift component, which correpsonds to the $L_1$ lift component in the von-K\'arm\'an and Sears notation. The term $\iint {x\omega_z \; dx dy}$ represents the first $x$-moment of the vorticity. The equation derived by Wu~\cite{Wu1981} is based on the principle that if the vorticity distribution over the entire flow field were known the force could be evaluated accurately. Although the lift terms may, for utility and convenience, be divided farther into $L_0$ and $L_2$ components, such division is to some extent arbitrary. According to this approach, the circulatory lift $L_c(t)$ is equal to the time rate of change of the first moment of the vorticity field. By applying the Taylor hypothesis, $dx = U_c dt$, one can transform the spatial derivative into a temporal one. In the unsteady thin airfoil terminology the circulatory lift component is only a portion of the total lift that acting on the flapping wing. Since at the beginning of the flapping cycle the lift is unknown we refer to the estimated lift component as an increment in the circulatory lift that is generated from the beginning of the cycle, thus equal to $\Delta L_c(t)$ and can be expressed as

\begin{equation}
\Delta L_c(t) = \rho U_{\infty}  \int U_c\zeta(t)dt
\label{e:Panda_Lift}
\end{equation}

In order to estimate the circulatory lift $\Delta L_c(t)$ from Eq.(\ref{e:Panda_Lift}) one needs to acquire information regarding the vorticity flux $\zeta (t)$ in the near wake

\begin{equation}
\zeta (t) = \int \omega_z(t) dy
\label{e:Panda_Lift_b}
\end{equation}

\begin{figure}[p]
\subfigure[]
{\includegraphics[width=1\textwidth] 
{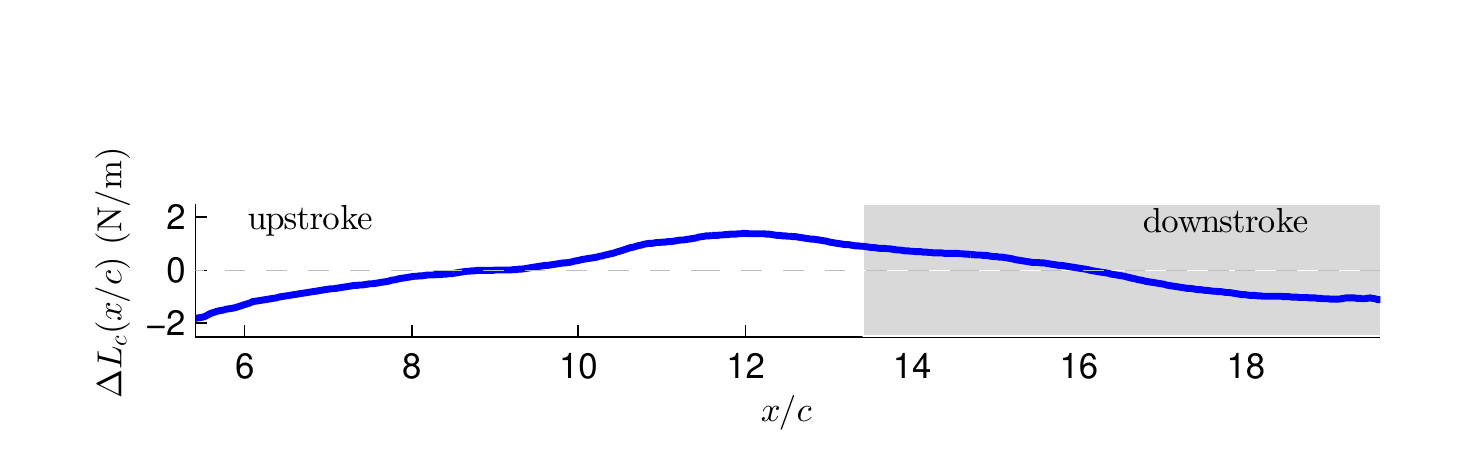}\label{f:Wake_Reconstruction_WB2a}}
\hspace{40pt}
\subfigure[]
{\includegraphics[width=1\textwidth] 
{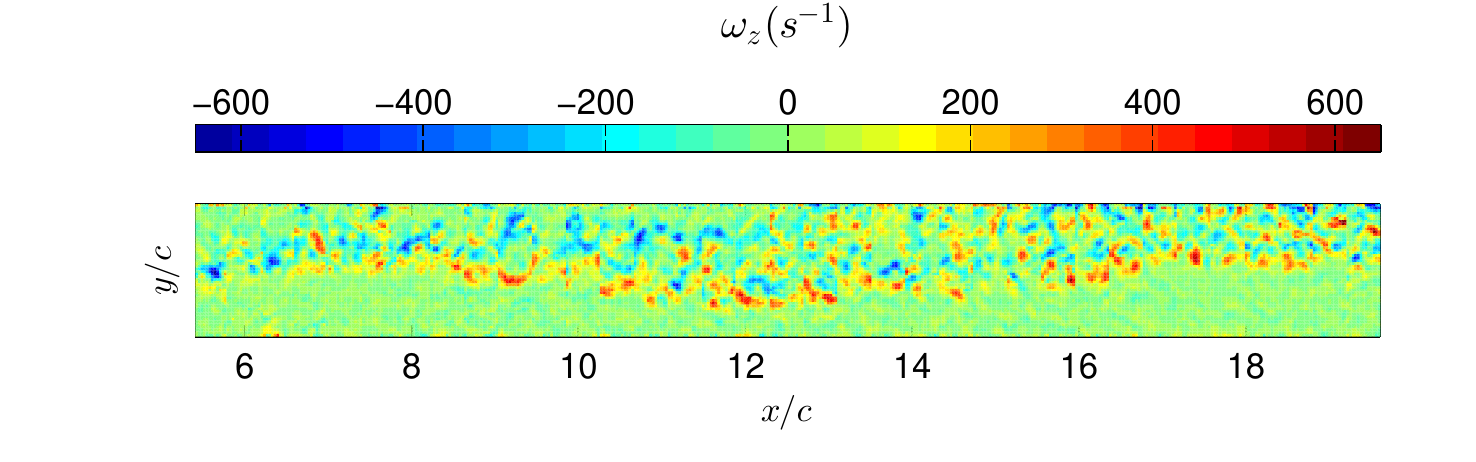}\label{f:Wake_Reconstruction_WB2c}}
\caption{{\bf Estimation of circulatory lift component based on wingbeat number 1.}
\\
\subref{f:Wake_Reconstruction_WB2a}~The circulatory lift was estimation based on eq.(\ref{e:Panda_Lift}). The grey area indicates downstroke flapping phase.
\subref{f:Wake_Reconstruction_WB2c}~Reconstruction of the starling's wake vorticity as thought the bird flies from right to left.}
\label{f:Wake_Reconstruction_WB2}
\end{figure}

\begin{figure}[p]
\subfigure[]
{\includegraphics[width=1\textwidth] 
{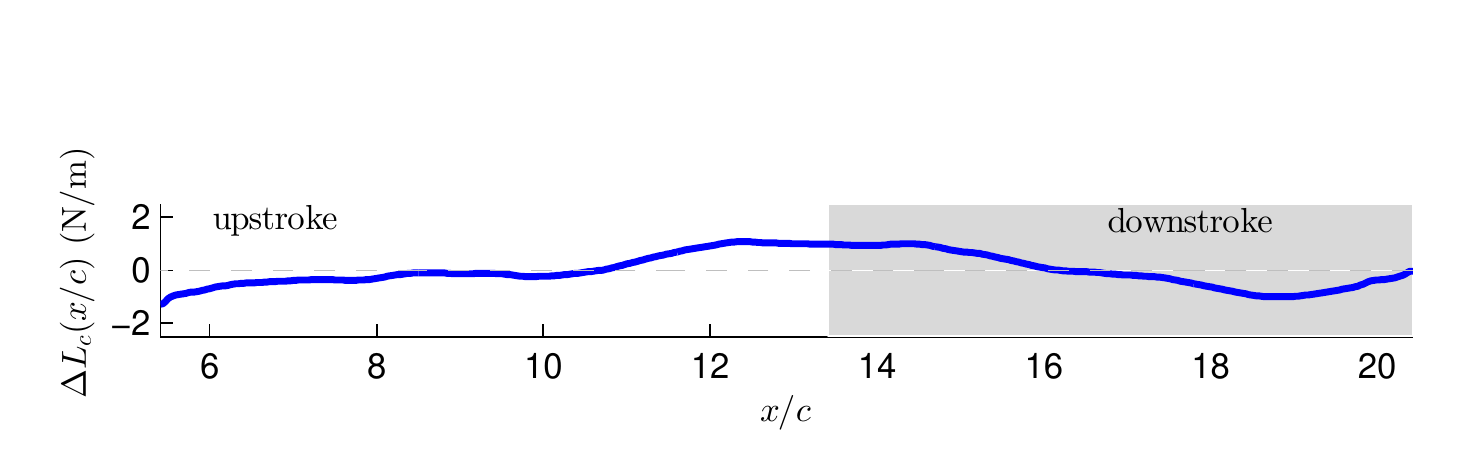}\label{f:Wake_Reconstruction_WB3a}}
\hspace{40pt}
\subfigure[]
{\includegraphics[width=1\textwidth] 
{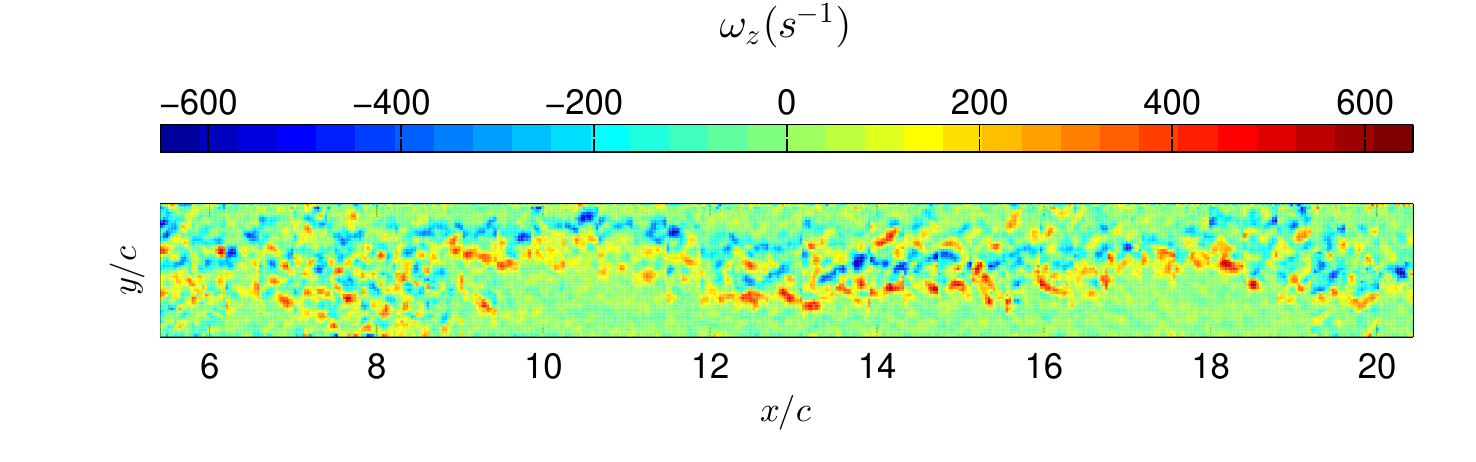}\label{f:Wake_Reconstruction_WB3c}}
\caption{{\bf Estimation of circulatory lift component based on wingbeat number 2.}
\\
\subref{f:Wake_Reconstruction_WB3a}~The circulatory lift was estimation based on eq.(\ref{e:Panda_Lift}). The grey area indicates downstroke flapping phase.
\subref{f:Wake_Reconstruction_WB3c}~Reconstruction of the starling's wake vorticity as thought the bird flies from right to left.}
\label{f:Wake_Reconstruction_WB3}
\end{figure}

\begin{figure}[p]
\subfigure[]
{\includegraphics[width=1\textwidth] 
{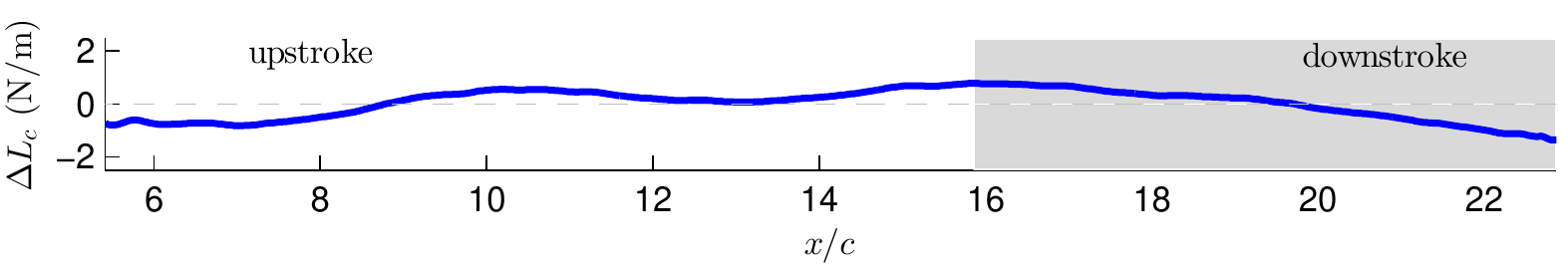}\label{f:Wake_Reconstruction_WB4a}}
\hspace{40pt}
\subfigure[]
{\includegraphics[width=1\textwidth] 
{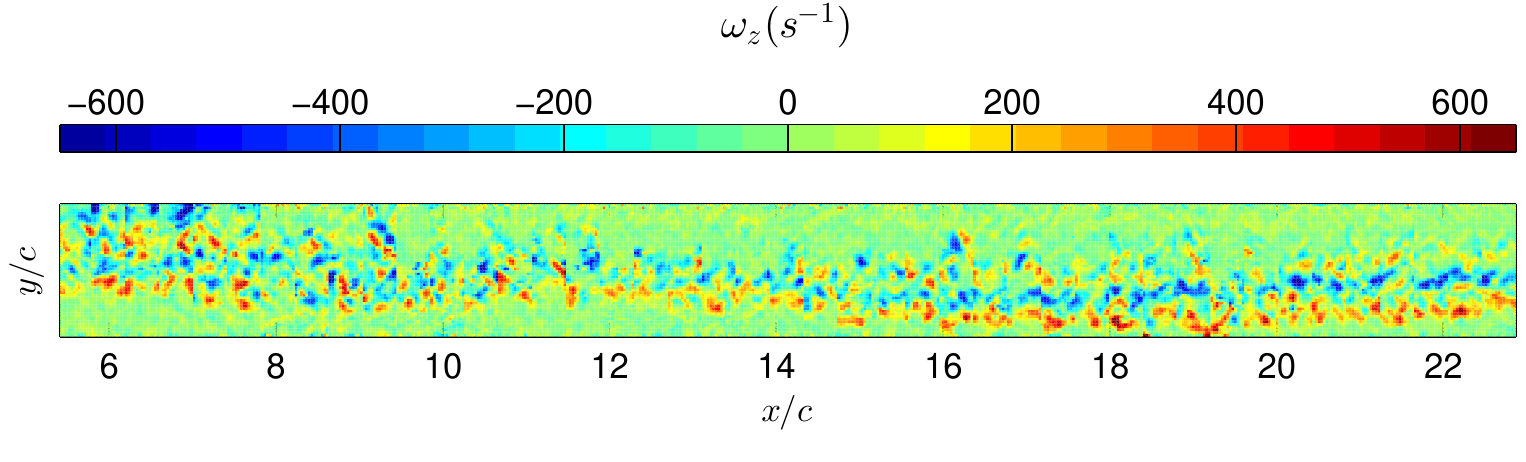}\label{f:Wake_Reconstruction_WB4c}}
\caption{{\bf Estimation of circulatory lift component based on wingbeat number 3.}
\\
\subref{f:Wake_Reconstruction_WB4a}~The circulatory lift was estimation based on eq.(\ref{e:Panda_Lift}). The grey area indicates downstroke flapping phase.
\subref{f:Wake_Reconstruction_WB4c}~Reconstruction of the starling's wake vorticity as thought the bird flies from right to left.}
\label{f:Wake_Reconstruction_WB4}
\end{figure}

The vorticity flux, defined by Eq.(\ref{e:Panda_Lift_b}), is estimated for each individual vector map as function of time. The calculated vorticity flux corresponds to the spanwise vorticity component and is integrated over a selected region in each vector map. The selected region covers the wake features that are observed in figures~\ref{f:Wake_Reconstruction_WB2c}, \ref{f:Wake_Reconstruction_WB3c}, and \ref{f:Wake_Reconstruction_WB4c}. Figure \ref{f:Wake_Reconstruction_WB2a} demonstrates the changes in the circulatory lift component as it evolves over time, calculated based on Eq.(\ref{e:Panda_Lift}) over a single wingbeat cycle. This calculation was performed for three different wingbeat cycles. The curve represents the cumulative lift over one wingbeat cycle, starting from right to left as the bird is moving from right to left in respect to the coordinate system. Overall, the lift accumulates positively during the downstroke phase whilst negative accumulation is depicted during the upstroke phase. One can deduce that the circulatory lift has a positive net effect during the downstroke phase, which is in agreement with former work~\cite{Hedenstrom2007, Warrick2005}. During the upstroke phase, it appears that the circulatory lift is decreasing; this implies that during this phase, the bird is losing energy through the unsteady mechanism. Furthermore, the presence of cumulative negative circulatory lift during the upstroke phase marks the energy that the bird has to invest in order to bring the wing back to the downstroke phase to generate lift, again. 

In the case of flapping flight of natural free-flyers, the wings' motion is extremely complicated and it comprises from a complex flapping motion (changes in effective angle of attack), wing deformation (the bird stretches or bending its wing) and substantial three dimensional motion~\cite{Brown1963}, thus resulting in a complex wake vorticity system~\cite{Kirchhefer2013}. As was shown earlier by the unsteady thin airfoil theory the vorticity shed into the wake continuously and affect the total circulation around a lifting surface~\cite{Theodorsen1935, KarmanSears1938}. Therefore, as demonstrated in figures~\ref{f:Wake_Reconstruction_WB2}, \ref{f:Wake_Reconstruction_WB3}, and \ref{f:Wake_Reconstruction_WB4}, the evolution of the lift over a single wingbeat cycle should be considered even for the case of power estimates where it is shown that while on average the lift should be equal to the bird's weight, the time dependent variations of the lift from this value might provide a plausible argument to a more efficient flight.

Pennyquick~\cite{Pennycuick1969} suggested to use the quasi-steady approach when applied to estimating lift in bird flight. In his work, power was estimated based on kinematic analysis of flying birds and some assumptions related to drag and lift. The lift was assumed to be equal to weight, as any body that is aloft and in equilibrium. Following this approach, Tucker~\cite{Tucker1973} revisited this argument and refined it to consider the flapping motion of the wings. Based on these works, Rayner~\cite{Rayner1979} proposed a mathematical model for lift estimation from the wake of a flying bird in various flight modes. This model is well supported in the literature and it is conceptually accepted that unsteady mechanisms are of minor importance since variation of wing pitch and circulation are not producing thrust~\cite{Rayner2001}. Our results, on the other hand, demonstrate that the unsteady mechanisms indeed play a significant role in the generation of lift. Whilst, thrust is not generated by lift, it is the energy that is required to keep the bird aloft that is impacted by the lift mechanism, e.g.: with less power required to generate lift, more power can be directed towards propulsion.

The argument that during the steady phase of flight the bird's weight must be balanced by an equal amount of lift~\cite{Pennycuick1969} is obviously valid for an average lift generated during a wingbeat cycle. In the current study the bird's weight was equal to $2$ N/m. Following this argument, for steady flapping flight, the starling is required to generate an equal amount of lift force. The three cycles show similar trends, where the lift variation over each wingbeat cycle is about $\pm 2$ N/m. As mentioned above, these lift values are the time variation of the circulatory lift component. Thus, in order to obtain the actual total lift force produced by the starling, one needs to add these fluctuations to the bird's weight that represents the averaged lift over a wingbeat cycle. By adding the lift fluctuations with the starling's weight we find that during the flapping cycle the starling can produce up to $4$ N/m of lift force, which is also equal to twice of its body weight. These maximum lift values are generated mostly during the downstroke to upstroke transition phase, as depicted in figure~\ref{f:Wake_Reconstruction_WB2c}, \ref{f:Wake_Reconstruction_WB3c}, and \ref{f:Wake_Reconstruction_WB4c}, which is in good agreement with Kirchhefer et al. 2013~\cite{Kirchhefer2013} who showed particular flow structures (termed: 'double branch features') occurring during the downstroke to upstroke transition. In addition, one can observe that the the minimum lift values are produced during the transition from upstroke to downstroke phase. The physical argument to this observation is that during the upstroke to downstroke transition the starling folds its wings, causing them to stop acting as lifting surfaces, and thus generates almost no lift. The aforesaid results imply that the usage of quasi-steady lift theory might be underestimating the lift that a bird is actually generates during flapping flight. 

\section*{Conclusions}
The objective of this study was to evaluate the effect of flapping motion on the generation of unsteady lift. Therefore, freely flying European starling's kinematics and near wake flow fields were acquired simultaneously at the AFAR facility. The bird's kinematics were measured with high speed imaging system, whilst the near wake was acquired with long duration time resolved PIV. 

To estimate the time dependent lift we have applied unsteady thin airfoil theory as developed by Theodorsen~\cite{Theodorsen1935} and von-K\'arm\'an and Sears~\cite{KarmanSears1938}. The theory addresses the various mechanisms that contribute to the time dependent lift components: the quasi-steady, added mass and wake-induced vorticity. Using these terms, we have estimated the lift from the wingbeat kinematics and demonstrated the contribution of each one of the terms to the total lift.

The theory assumes a planar wake and a trailing-edge Kutta condition. Thus it excludes wake roll-up, convection of large separations over the wing section, boundary-layer separation, large laminar separation bubbles, leading-edge and trailing-edge vortices, three-dimensional effects and so forth. However, these effects are significant at low Reynolds numbers and thus lift estimated with the unsteady thin airfoil theory only partial explains the complex flow physics.

The aerodynamic forces estimated from the wake flow field measurements as acquired with PIV provide a more accurate estimation of the aerodynamic forces. The equations that provide the basis to determine the aerodynamic forces were derived by Wu~\cite{Wu1981}. As discussed by Panda~\cite{Panda1994} only the circulatory lift component can be estimated from the wake measurements. In this work, three wingbeat cycles were analysed. The near wake behind the bird's wing was spatially reconstructed from the temporal data. The reconstructed wake field is spanning over 20 chord lengths downstream and depicts the flow features in the wake. The circulatory lift component over the wingbeat cycle follows the flow features as shown by the reconstructed wake. 

The evolution of the circulatory lift presents a negative contribution during the upstroke phase, whilst during the downstroke phase a positive contribution is observed. In addition, we observed that the downstroke phase is shorter, compared to the upstroke phase. This asymmetrical pattern is essential for the production of the high lift impulse that supports the bird's weight. We conclude that the bird prefer to generate high lift values with a short downstroke phase than moderate lift values with a longer downstroke phase. 

The variation of the lift over the wingbeat cycle emphasizes its contribution to the total lift and its role in power estimations. We suggest that the time dependant circulatory lift component cannot be assumed negligible and should be considered when estimating lift or power of birds in flapping motion.

\bibliographystyle{plos2009.bst}
\bibliography{references}

\end{document}